\documentclass[11pt]{article}

\usepackage[margin=1.15in]{geometry}
\usepackage{amsmath}
\usepackage{amssymb}
\usepackage{amsthm}
\usepackage[round,authoryear]{natbib}
\usepackage{booktabs}
\usepackage{microtype}
\usepackage{enumitem}
\usepackage{hyperref}
\hypersetup{colorlinks=true,linkcolor=blue,citecolor=blue,urlcolor=blue}

\theoremstyle{plain}
\newtheorem*{namedthminner}{\namedthmname}
\newenvironment{nthm}[1]{\def\namedthmname{#1}\begin{namedthminner}}{\end{namedthminner}}
\theoremstyle{definition}
\newtheorem*{nameddefinner}{\nameddefname}
\newenvironment{ndef}[1]{\def\nameddefname{#1}\begin{nameddefinner}}{\end{nameddefinner}}

\newcommand{\prm}{\texorpdfstring{$'$}{'}}

\title{Removable Defects: The Economics and Limits of Deliberate
  Deficiency\thanks{Extended notes, the research backlog, and per-claim
  literature verdicts are maintained at
  \url{https://github.com/macrokit/removable-defects}.}}
\author{Cheng Qian}
\date{July 13, 2026}

\begin{document}
\maketitle

\begin{abstract}
A specialist tolerates blind spots that a generalist does not. Usually this is
treated as a cost to be minimized. We treat it as a design variable: a
deficiency can be \emph{kept} because it pays --- it concentrates capability and
raises output where the specialist is strong --- and \emph{removed on demand} in the
rare situation where it would be fatal, by routing that situation to a
compensation channel instead of acting. We give three results. First, an
\emph{advantage condition} under which keeping the deficiency is a computable
economic position rather than a character trait; structurally it is the
Ehrlich--Becker market-vs-self-insurance margin applied to a competence gap,
with the detector as a Townsend costly-state-verification technology. Second,
a two-sided characterization of when removal is possible. The obstruction is
captured by a \emph{coupling lemma}: when the deficiency is a coarsening of
perception, the rate at which the specialist's feature does good equals the
rate at which it does harm --- one number prices both --- so no switch, however perfect, can separate
benefit from harm. This yields a converse (a confounded detector earns zero
premium --- and any within-defect policy insisting on positive premium is
driven, under multiplicative dynamics, to long-run growth of negative
infinity) and a matching achievability result (a detector placed
\emph{outside} the deficiency earns an explicit positive premium under the
advantage condition). Together, over structured uncertainty classes and with
severity capped or the miss rate scaling as $O(1/L)$: \textbf{a defect is
profitably removable iff the
detector-relevant distinction survives the restriction and the advantage
condition holds} --- and a single value formula covers every detector class
(the premium is the support function of the class's ROC set at an economic
price vector, with removability a coefficient, not a yes/no). Third, a
distinction between \emph{observation} and \emph{capacity} defects that is invisible to
the average-case literature but sharp here --- they differ exactly on whether
access to the deployment distribution rescues them, and the gap between the
two decomposes exactly: it is cross-leak plus a closure deficit, and per-task
randomization buys back the latter, never the former. The characterization is
constructive end to end --- the detector can be \emph{learned} from declared fatal
categories at a one-time training bill linear in the loss severity, with the
advantage condition surviving learning. The results synthesize Chow's reject
option, Kelly growth under ruin, and selective prediction; the residual
contributions are the converse as a worst-case impossibility, the detector's
per-event miss rate (not the event frequency) as the binding constraint, and
a frequency-free form of the advantage condition.
\end{abstract}

\section{Introduction}\label{sec:intro}

A deliberately narrow system guarded by a tripwire and an escalation path is
one of engineering's most common patterns, and the literatures that study its
parts are mature. Chow's reject option prices the tripwire's threshold;
selective prediction and learning-to-defer train it; out-of-distribution
detection asks how well it fires where the data runs out; AI control demands
a monitor outside the untrusted system \citep{greenblatt2024}. All of them ask how \emph{good} the
detector is. \textbf{None asks where the detector sits relative to the restriction
that created the blind spot it guards} --- whether the narrowing that makes
the specialist cheap also reaches the tripwire. This paper shows that
placement is the whole game.

The central fact is a two-line lemma with a long shadow. Model the
specialist's deficiency as a coarsening of perception, and force the detector
to work through that same coarsened view. Then for any pair of task ensembles
the coarsening cannot tell apart --- one safe and profitable, one fatal --- \textbf{the
detector's gain-capture rate equals its fatal-miss rate: one number prices
both.} Every unit of specialization gain let through carries an equal unit
of fatal exposure, so benefit and harm cannot be decoupled by anything
downstream of the coarsening; a switch that disables the harm disables the
benefit in the same motion, however perfect the switch. From this coupling
the paper's two-sided characterization follows: \textbf{a defect is profitably
removable iff the detector-relevant distinction survives the restriction ---
iff the detector can be placed outside the defect --- and an explicit advantage
condition holds.}

The general object behind the story is a feature whose value depends on
context: harmful in situation $A$, beneficial in situation $B$. Always-on
pays $A$'s harm; always-off --- the generalist who never commits --- forfeits
$B$'s benefit; \emph{controlled} (enable in $B$, disable in $A$) dominates both,
if achievable. Control needs a switch, which is usually cheap, and a detector
that tells you which situation you are in, which is the binding constraint: a
flawless switch you cannot time is worthless. The economically sharp case is
asymmetric --- $B$ common and mildly good, $A$ rare but \emph{fatal} --- and there the
feature is kept enabled by default and switched off only on detection, by
routing the rare fatal case to a compensation channel (buy the missing work,
defer to a principal, rent judgment). We call such a deliberately-kept,
on-demand-correctable deficiency a \textbf{removable defect}.

Two questions organize the paper. \emph{When is keeping the defect worth it?} --- an
economic question, answered by the advantage condition (\S\ref{sec:advantage}). \emph{When can it
actually be removed?} --- a structural question, answered by the coupling lemma
and its consequences (\S\ref{sec:removability}--\S\ref{sec:taxonomy}): removal can be impossible even with a perfect
switch, and the obstruction is precisely where the detector sits.

\subsection*{Contributions}

\begin{enumerate}
\item \textbf{The coupling lemma, and the converse it forces} (\S\ref{sec:removability}; appendix Lemma~1
   and Theorem~1, Appendices \ref{app:coupling}--\ref{app:thm1}). Inside an observation defect, gain-capture rate $=$ fatal-miss
   rate --- benefit and harm are one scalar. Consequently a confounded detector
   earns zero specialization premium against an adversarial task distribution
   --- \emph{even with transductive access to the deployment distribution} --- and
   under multiplicative (Kelly) dynamics any positive premium forces
   log-growth to $-\infty$. To our knowledge the first worst-case,
   decision-theoretic formalization of detector placement --- the principle AI
   control and safety engineering hold as an axiom, given formal backing.
\item \textbf{The two-sided iff, priced} (\S\ref{sec:advantage}, \S\ref{sec:removability}; appendix Theorem~A,
   Appendix~\ref{app:thmA}). A detector outside
   the defect earns an explicit positive premium under an \textbf{advantage
   condition} that is the Ehrlich--Becker insurance margin restated for a
   competence gap, with the detector a Townsend verification cost. Placement
   supplies the possibility; the advantage condition supplies the profit. The
   condition is frequency-free: the ruin condition is written on the
   detector's per-event miss rate, never on the fatal-event frequency that
   history cannot estimate.
\item \textbf{A taxonomy with an exact price} (\S\ref{sec:taxonomy}; Appendices
   \ref{app:thm1prime}--\ref{app:thmM}). \emph{Observation}
   defects (the information is destroyed) and \emph{capacity} defects (the
   information survives; the class cannot use it) differ exactly on whether
   deployment knowledge rescues them --- yet one value formula prices both:
   premium $=$ support function of the detector class's ROC set at an
   economic price vector, with removability a coefficient, not a yes/no. The
   gap between menu and machine decomposes into two named obstructions with
   exact formulas, a router bound that makes the theory recurse on itself,
   and a least-favorable-mixture duality: \textbf{a coin buys back the gluing
   failure, never the cross-leak.}
\item \textbf{End-to-end learnability} (\S\ref{sec:detector}; Appendix~\ref{app:thmE}). The detector can be
   selected and certified from stratified samples of declared fatal
   categories: a one-time training bill \emph{linear} in the loss severity plus
   $O(\log L)$ per-period rent, both growing slower than the loss they
   insure --- the advantage condition survives learning, and default-deny
   gating makes the certificate fail-safe against cover incompleteness.
\end{enumerate}

\section{The model}\label{sec:model}

Tasks arrive one per period from a distribution $P$ on a space $X$. The
specialist has a \textbf{competence set} $C \subseteq X$ it handles well; the
complement is the defect. A \textbf{fatal set} $F$ marks tasks where acting without
competence is catastrophic; the danger zone is the \textbf{fatal exposure}
$E := F \setminus C$, with mass $\varepsilon := P(E)$, assumed small. Per task
the agent either \emph{acts} or \emph{escalates} to the compensation channel:

\begin{center}
\begin{tabular}{lcc}
\toprule
 & $x \in C$ & $x \in E$ \\
\midrule
\textbf{act} & $+g$ & $-L$ \\
\textbf{escalate} & $-p$ & $-p$ \\
\bottomrule
\end{tabular}
\end{center}

\noindent
with $g$ the specialization gain, $L \gg g$ the fatal loss, $p$ the price of
compensation. A \textbf{detector} $d : X \to \{\text{act}, \text{escalate}\}$ gates
the choice, at running cost $c_d$, with two error rates: the \textbf{miss rate}
$\mu = P(d{=}\text{act} \mid E)$ (acting on a fatal task --- the dangerous
error) and the \textbf{false-alarm rate} $\alpha = P(d{=}\text{escalate} \mid C)$
(escalating work the specialist could have done --- the self-inflicted tax). The
detector never \emph{solves} the escalated task; it answers one binary question,
``is this mine?'', and routes.

Expected per-period growth is
$$G = (1-\alpha)\,g\,P(C) - c_d - p\,\big[\alpha P(C) + (\text{escalations outside } C)\big] - \varepsilon\mu L - (\text{benign terms}).$$

Two structural facts. The fatal term $\varepsilon\mu L$ is a product of three
factors; the architecture attacks $\mu$ (the detector) and caps $L$ (bounded
blast radius --- a spend ceiling), and deliberately does \emph{not} lean on
estimating $\varepsilon$ well, because tail frequencies are exactly what
history gets wrong. And the generalist alternative is not free: it pays a
carrying cost $c_{\text{gen}}$ per period for rarely-used breadth and earns
$g_{\text{gen}}$ (typically $< g$) on
$C$, since breadth dilutes depth.

\section{The advantage condition}\label{sec:advantage}

\begin{nthm}{Proposition 1}
The specialist-with-detector beats the generalist iff the
specialization gain exceeds the cost of insuring the tail:
$$\big[(1-\alpha)g - g_{\text{gen}}\big] P(C) + c_{\text{gen}} \;>\; c_d + p\,\mathbb{E}[\text{escalations}] + \varepsilon\mu L.$$
\end{nthm}

Informally: \emph{specialization gain per period $>$ fatal-event frequency $\times$ price of
compensation}, once the detector's rent and false alarms are charged. When
fatal events are rare, the deliberate deficiency wins, and every term is
measurable in an instrumented environment --- which is what makes ``should I be
deficient?'' a computable question rather than a matter of temperament.

This is not new economics, and we do not claim it is. Structurally it is the
Ehrlich--Becker \citeyearpar{ehrlich1972} substitution margin between \emph{market insurance} (the
per-event compensation channel) and \emph{self-insurance} (the generalist's carried
breadth), with the detector playing the role of Townsend's \citeyearpar{townsend1979} costly state
verification --- pay to verify, and only in the bad state. The carrying-cost-
versus-per-event framing is the standard make-or-buy / capacity-versus-spot
tradeoff \citep{williamson1985,vanmieghem2003}. Our contribution here is narrow and
we state it as such: the \emph{insured object} is a deliberately maintained
competence gap rather than a portfolio position, and --- see \S\ref{sec:removability} --- the binding
constraint turns out to be the detector's reliability, not the frequency the
insurance is priced against.

\section{The detector: what it costs and where it must sit}\label{sec:detector}

\textbf{The optimal rule is Chow's.} Because $G$ is linear in $(\mu, \alpha)$, the
loss-minimizing escalate-versus-act rule is a likelihood-ratio test whose
threshold is the ratio of what a false alarm wastes to what a miss costs. This
is exactly Chow's reject option \citep{chow1957,chow1970}; we cite it and keep only one
design consequence: \textbf{the threshold is an economic quantity, not an
engineering constant.} As stakes $L$ rise, the same detector must slide along
its ROC curve toward paranoia; a system whose escalation threshold cannot be
re-priced is mis-built. One caveat we will need \citep{cortes2016}: with a \emph{restricted} actor, the optimal rejection region is provably not
a threshold on the actor's own confidence --- the detector must be a separate
function. This foreshadows \S\ref{sec:removability}.

\textbf{Detection is cheap --- and that is load-bearing.} Detecting whether a task is
out of scope is a binary hypothesis test; by the Chernoff--Stein bound
\citep[Thm.~11.8.3]{coverthomas2006} its cost
scales as $\log(1/\mu)$ in the miss rate. Producing the \emph{solution} to an
out-of-scope task is far more expensive. We will see in \S\ref{sec:removability} that survival
against fatal events of severity $L$ forces the miss rate down to $O(1/L)$,
which by the hypothesis-testing bound costs only $O(\log L)$ in detector rent.
The architecture insures against loss $L$ at logarithmic price --- this is why
it is affordable at all. (Putting detection and production in a single cost
model --- rather than the mismatched units of samples and compute --- is possible
via doubly-efficient interactive proofs and property testing; we regard this
as a strengthening, not a load-bearing step, and do not pursue it here.)

The shape survives when the detector must be
\emph{learned} rather than assumed: certifying a near-zero miss rate from data is
a one-sided rare-event problem, costing a \textbf{one-time training bill linear in
$L$} in labeled fatal examples (Appendix~\ref{app:thmE}) --- amortized over deployment,
both bills grow slower than the per-event loss they insure against, so the
advantage condition survives learning. A side consequence worth naming: the
scarce input is labeled catastrophes, which a well-run system stops producing
--- it escalates them --- so detector training data must come from history,
simulation, or other agents' failures.

\section{Removability: the central result}\label{sec:removability}

\subsection{Profitable removability}\label{sec:profitable}

We first correct a definition. It is tempting to call a defect \emph{removable} if
some policy keeps growth bounded below. This is vacuous: the constant
always-escalate detector lies in every detector class and bounds $G \geq -p$
regardless of the defect. Survival is trivially available; it is not the
question. The question is \emph{paid-for} survival:

\begin{ndef}{Definition}
The \textbf{specialization premium} of a detector is its worst-case
growth --- over an uncertainty class $\mathcal{P}$ of task distributions whose
fatal mass is capped by a declared budget $\bar\varepsilon$ --- \emph{above the
always-escalate baseline}. A defect is \textbf{profitably removable} on
$\mathcal{P}$ if some \emph{within-defect} policy --- detector drawn from the
restricted class the defect induces --- earns strictly positive premium: it
retains part of the specialization gain while avoiding ruin.
\end{ndef}

\subsection{The coupling lemma}\label{sec:coupling}

Formalize the defect as a restriction on what the agent can perceive: an
\textbf{observation defect} forces the detector to be a function of a coarsened
view $\varphi(x)$ rather than of $x$ itself. (This covers pruned, quantized,
or attention-limited perception.)

\begin{nthm}{Lemma}[coupling]
Let the detector factor through $\varphi$. For any pair
of task ensembles --- one safe, one fatal --- that $\varphi$ cannot tell apart,
the detector's gain-capture rate on the safe ensemble equals its fatal-miss
rate on the fatal ensemble.
\end{nthm}

The proof is two lines (Doob--Dynkin factorization plus equality of
pushforwards; Appendix~\ref{app:coupling}, Lemma~1). The content is in what it says: \textbf{inside an
observation defect the miss rate \emph{is} the capture rate.} Every unit of
specialization gain the detector lets through carries an equal unit of fatal
exposure --- the two rates are literally the same number. This is the
common-cause-failure principle of reliability engineering \citep{eckhardt1985,littlewood2012} upgraded from a correlation model to an identity. A
switch that disables the harm disables the benefit in the same motion.

\subsection{The converse}\label{sec:converse}

\begin{nthm}{Theorem 1}[converse]
Against an adversarial choice of task distribution
within the fatal-mass budget, a detector confined inside an observation defect
earns zero specialization premium once $L$ exceeds an explicit threshold --- and
this holds even if the detector is chosen after seeing the deployment
distribution. Under multiplicative (Kelly) dynamics with true ruin, any
positive premium forces expected log-growth to $-\infty$.
\end{nthm}

The mechanism is the coupling lemma: positive premium requires acting on the
confounded family at some positive rate $q > 0$, which by the lemma is also
the fatal-miss rate, and the missed-fatal term $-\varepsilon q L$ overwhelms
the gain as $L$ grows. An exact value formula (Appendix~\ref{app:thm1prime}, Theorem~1$'$) refines
``zero'' to a coefficient: the surviving premium fraction is the \textbf{residual
separability} $\sigma_0$ --- the share of safe mass the coarsening still places
at zero fatal risk. So removability is not a yes/no but a number: \emph{a defect is
profitably removable exactly to the extent that the detector's distinction
survives the restriction.}

That transductive access does not help is worth pausing on. Seeing the
deployment distribution tells you \emph{that} fatal mass is present; it never tells
you \emph{which task} is fatal, because $\varphi$ destroyed that distinction
per-task, before any distributional knowledge can act. Optimal play here is a
Chow rule on the coarsened posterior --- and at the confounded points that
posterior is uninformative, forcing all-or-nothing behavior. Chow's rule
executes the collapse; it cannot prevent it. The failure is not in the
decision rule but in what reaches it.

\subsection{Achievability, and the iff}\label{sec:achievability}

\begin{nthm}{Theorem A}[achievability]
A detector placed outside the defect ---
achieving miss rate $\leq \delta$ and false-alarm rate $\leq \alpha_0$
uniformly over a structured class with competence floor $\underline c$ and
fatal budget $\bar\varepsilon$ --- earns worst-case premium bounded below by the
explicit constant
$$\Pi^* = \underline c\,(1-\alpha_0)(g+p) - \bar\varepsilon\,\delta\,(L-p) - c_d,$$
and this guarantee is positive ($\Pi^* > 0$) --- so removal is profitable ---
iff the advantage
condition $\underline c(1-\alpha_0)(g+p) > \bar\varepsilon\delta(L-p) + c_d$
holds.
\end{nthm}

The premium reduces to three per-unit prices --- captured competence at $g+p$,
leaked fatal at $L-p$, flat rent --- and Theorem A is just Proposition 1
re-derived with the detector's real operating characteristics in place of the
abstract ``price of compensation.'' Existence of such an out-of-defect detector
is exactly a class-uniform coverage guarantee; it is achievable for structured
uncertainty classes (finite-VC families, divergence balls) and, by the
coupling lemma, \emph{impossible} inside an observation defect. Achievability
inherits that structural restriction --- no more, no less --- and the existence
assumption is ultimately discharged constructively: Appendix~\ref{app:thmE} selects the
detector from data within that structure, with certified rates.

The two theorems meet at the large-loss boundary. With $\delta$ fixed,
$\Pi^* \to -\infty$ as $L \to \infty$ --- a detector of any fixed miss rate is
eventually ruined, matching the converse. Survival against arbitrary $L$
requires either sharpening the detector to $\delta = O(1/L)$ (rent only
$O(\log L)$, per \S\ref{sec:detector}) or capping the blast radius. At the ideal operating point
the bound recovers the converse's separation witness exactly, so the two
results are two ends of one continuum.

\begin{nthm}{Corollary}[the iff]
A defect is profitably removable iff a detector with
small miss and false-alarm rates exists outside it --- iff the detector-relevant
distinction survives the restriction --- and the advantage condition holds.
\end{nthm}

Necessity is the converse (inside the defect the distinction is gone);
sufficiency is achievability (outside it, an explicit positive premium).
\textbf{Placement supplies the possibility; the advantage condition supplies the
profit.}

\section{Two kinds of defect}\label{sec:taxonomy}

The restriction in \S\ref{sec:removability} coarsened \emph{perception}. A different restriction confines
the detector to a limited \emph{function class} $H$ while perception stays full --- a
\textbf{capacity defect}. The two behave differently, and the difference is the
paper's cleanest structural claim.

For a capacity defect the premium collapses in the \textbf{inductive} regime --- when
the detector is committed before the deployment distribution is known
($\sup_d \inf_P$) --- but is fully recovered in the \textbf{transductive} regime, when
the detector may be chosen knowing the distribution ($\inf_P \sup_d$)
(Appendix~\ref{app:thm2}, Theorem~2). The inductive-versus-transductive gap is exactly the
mechanism by which \citet{goldwasser2020} escape
impossibility using unlabeled deployment data. For an observation defect, by
contrast, the converse holds \emph{even transductively} (\S\ref{sec:converse}): the gap is zero.

So the two defect types are provably distinct along the transduction axis ---
an observation defect destroys the distinction per task, and no amount of
distributional knowledge rebuilds it, while a capacity defect still contains
the right detector; it merely cannot be identified until deployment is seen.
This is the switch/detector
distinction of \S\ref{sec:intro} made precise: a capacity defect is ``the switch is confined
but the context is in principle visible''; an observation defect is ``the
context itself is unresolvable to you, and switch quality is irrelevant.'' The
average-case selective-prediction literature, which fixes a distribution, does
not see this split; the worst-case framing makes it unavoidable.

The split is sharper than it first appears, because it is \emph{not} a split in
how removability is priced. A single value formula covers both defect types
(Appendix~\ref{app:thm2prime}, Theorem~2$'$): for \emph{any} detector class, the worst-case premium is
the support function of the class's ROC set --- its achievable (capture, leak)
pairs against the confusable ensembles --- evaluated at the economic price
vector (captured competence at $g+p$, leaked fatal at $L-p$), and the
removability coefficient is the class's \textbf{zero-leak capture capacity}, of
which \S\ref{sec:converse}'s $\sigma_0$ is the $\sigma$-algebra special case. What genuinely
separates the two types is \textbf{joint realizability}: on any single confusable
direction the inductive and transductive values coincide for every class, so
the transduction gap exists only when the adversary has several directions,
each resolvable by some member of the class but no member resolving all. An
observation defect fails per direction; a capacity defect fails only jointly.
One formula for the price; two escape routes.

The joint-realizability deficit is itself characterizable (Appendix~\ref{app:deficit}). It
decomposes into exactly two obstructions --- \emph{cross-leak} (one direction's
solution acts on another's fatal mass) and \emph{non-closure} (the class cannot
glue its local solutions) --- with exact formulas where each acts alone: for
confidence-threshold detectors the deficit is a ``global paranoia'' tax (the
safe mass trapped between each direction's fatal ceiling and the worst
direction's), and for halfspaces it is precisely non-separability of the safe
and fatal point clouds. In general the deficit is bounded by the confusion
mass of a \emph{direction router}, priced at the same two prices --- and since a
router is itself a detector (of directions rather than of fatality), the
question recurses. The recursion is the observation/capacity split in its
sharpest form: for a capacity defect it terminates in a cheap router,
obeying the same $O(\log L)$ detector economics; for an observation defect
the router would need exactly the information the coarsening destroyed, and
the recursion halts at the coupling lemma with nothing to buy. When the two
obstructions mix, a duality resolves them (Appendix~\ref{app:thmM}): allowing the agent
per-task randomization convexifies its options, and the many-direction game
collapses to a single-direction problem at a \emph{least-favorable mixture} of
directions --- whereupon the transduction gap splits additively into
irreducible cross-leak plus the price of determinism. \textbf{A coin buys back the
gluing failure, never the cross-leak.}

\section{A worked example: Agent World}\label{sec:agentworld}

The theory grew out of an engineering problem, and the engineering carries an
instantiation of it. \emph{Agent World} is a protocol for autonomous agents that
act on a principal's behalf: each agent is a keypair, a signed manifest, and an
inbox. The instantiation is at the design and protocol level --- a manifest field,
its schema, and a conformance rule, merged to the project's mainline with its
chain test passing --- not a deployment with measured data. We describe it as a
demonstration that the paper's objects have a natural protocol footprint, not
as empirical confirmation.

\textbf{The defect is a closed list.} A manifest declares the agent's capabilities
as a \emph{positive, closed} list of typed task classes. Everything not listed is a
deliberate deficiency, and silence is the honest way to express it --- the
competence set $C$ of \S\ref{sec:model}, written down. A narrow agent with a high capability
score on one class is the specialist the advantage condition rewards; a diluted
generalist is what it warns against.

\textbf{The compensation route is declared, the detector is not.} A manifest may
carry one optional field:

\begin{verbatim}
"onOutOfScope": "escalate:market"    // or "escalate:owner" | "decline"
\end{verbatim}

This names the route the agent takes when it meets work outside its competence:
post the missing task to the market, defer to its principal, or refuse.
Crucially, the field declares \emph{that} a route exists --- never the detector that
fires it. What recognizes ``this task is out of my competence and high-stakes''
is internals, out of the protocol's scope by construction. This is exactly the
paper's division of labor: the switch and its route are public and cheap to
state; the detector is the agent's own always-on burden.

\textbf{Placement is enforced, weakly but measurably.} No hub can verify that a
detector actually sits outside the defect. The protocol does two enforceable
things instead. It rejects an \emph{internally inconsistent} declaration --- a
manifest claiming \texttt{escalate:market} while its mandate lacks the right to post
tasks is read as \texttt{decline}, since it has declared a route it cannot take
(a conformance requirement, checked by a chain test that also rejects unknown
routes). And it makes the rest a \emph{measurable divergence}: declaring a route,
like declaring a goal frame, turns silent failure into an observable gap
between declared route and behavior. The advantage condition's terms are, by
design, quantities the hub can read --- capability score against market prices on
one side, observed escalation spend on the other --- so ``should this agent be
deficient?'' is meant to be computed, not judged.

\textbf{The two-sided theorem, from both sides at once.} Agent World names the
paper's central rule as its \emph{one load-bearing constraint}: ``the deficiency
must never include the detector --- an agent may be deficient in what it can do,
never in noticing what it cannot.'' We do not claim independent discovery --- the
design and this paper share an origin. What we can honestly claim is the
order: the constraint was articulated as an engineering necessity \emph{before} any
theorem existed, and the theorem then showed the necessity was mathematical,
not stylistic. And the two failure accounts line up as one fact seen twice.
The paper's converse is \emph{ex ante}: a confounded detector cannot earn premium,
full stop. Agent World's ledger supplies the \emph{ex post} image: an agent that
declares a route but whose detector is in fact blind will act out of
competence, overclaim, burn its stake, and go broke --- dissipation is the
punishment the market administers to the unremovable defect. Theorem 1 says it
was never removable; the ledger says why that costs.

\textbf{Removal in the strongest sense.} One of the declared routes is to \emph{buy a
capability module} --- after which the missing competence is no longer missing.
This is the permanent end of the removability spectrum: the defect does not
merely get compensated for one event, it \emph{becomes a capability}, and the word
``removable'' is then literal. The sealed-frame agent is the opposite extreme ---
a permanent deficiency that is never lifted, where removal means the guardian
machinery compensating the seal's one fatal case. That a single manifest field
spans permanent, per-event, and interpretive removal is the concrete form of an
open question we flag in \S\ref{sec:limitations}: whether ``removable'' is one concept or three.

\section{Related work}\label{sec:related}

Each ingredient has a home in an existing literature; we claim the coupling,
the converse, the taxonomy, and the synthesis. Per-claim verdicts with
sources are maintained in the project repository; here we position only.

\begin{itemize}
\item \textbf{Reject option / selective prediction.} The optimal escalate-versus-act
  rule (\S\ref{sec:detector}) is Chow's \citep{chow1957,chow1970}; \citet{cortes2016} show a
  restricted actor forces a separate rejector, an architecture the taxonomy
  already names \citep{hendrickx2024}. None of it provides a growth
  statement, and none asks where the rejector sits relative to the actor's
  restriction. The empirical finding that a rejector \emph{sharing} the
  actor's representation can beat an independent one \citep{geifman2019} is
  not a counterexample: that is average-case selective risk on the deployment
  distribution, while our placement claim is a worst-case guarantee against
  declared fatal events --- exactly the regime where Lemma 1 prices the shared
  representation at zero.
\item \textbf{Impossibility.} Adversarial-distribution impossibility for restricted
  detector classes is standard technology (\citealt{fang2022}; \citealt{ulmer2021}
  for same-representation failures). Unstated anywhere we could find: the
  \emph{coupling} --- one restriction defining both competence and blindness --- with
  an unbounded-loss consequence and a positive converse. The AI-control
  literature \citep{greenblatt2024} holds ``monitor outside the untrusted
  system'' as an axiom with no formal backing; this is the backing.
\item \textbf{Growth and insurance.} \citet{kelly1956} and \citet{breiman1961} own ruin
  exclusion; \citet{peters2015} and \citet{spitznagel2021} the licensing half;
  \citet{ehrlich1972} plus \citet{townsend1979} the advantage condition's
  structure. Our residual: the ruin condition written on the detector's
  \emph{miss rate}, not the event frequency.
\item \textbf{Common-cause failure.} \citet{eckhardt1985} and \citet{littlewood2012}
  give ``a monitor sharing the system's failure mode gives no protection'' as
  an average-case correlation model; the coupling lemma is its worst-case,
  identity-strength form.
\item \textbf{Minimax testing.} The duality behind the obstruction split (\S\ref{sec:taxonomy}) is
  classical --- \citet{wald1945,wald1950}; \citet{huberstrassen1973} for the collapse to a
  least favorable pair --- and randomization realizing the ROC convex hull is
  the Neyman--Pearson lemma \citep{neymanpearson1933} plus \citet{provost2001}. We claim the
  application only, and distinguish \citet{managoli2025}, who pair robust
  testing with abstention at asymptotic exponents over unrestricted
  detectors.
\item \textbf{Learning the detector.} Theorem E is \citet{vapnik1971} plus the
  BEHW \citeyearpar{blumer1989} version-space bound. Theorem E$'$'s problem is Kalai--Kanade--Mansour's
  positive-reliable learning \citep{kalai2012} and its setup is Neyman--Pearson
  classification at $\alpha = 0$ \citep{scottnowak2005}; neither contains the
  zero-empirical-miss fast-rate certificate nor the linear-versus-quadratic-
  in-severity bill --- itself an elementary corollary of the realizable/
  agnostic rate dichotomy. Closest neighbor: \citet{casacuberta2025},
  group-wise reliable abstention at agnostic rates.
\end{itemize}

\section{Limitations and open questions}\label{sec:limitations}

We are honest about the boundary of what is proved.

\begin{itemize}
\item \textbf{The learned detector's certificate is only as good as its cover.} The
  end-to-end result (Appendix~\ref{app:thmE}) trains and certifies the detector from
  stratified samples of \emph{declared} fatal categories, and default-deny gating
  makes it fail-safe against categories left out of the cover --- an unknown
  fatal category is escalated, not missed. What remains exposed is
  \textbf{mis-declaration}: fatal mass inside a declared-safe cell, where the gate
  admits and the samples mislead. That is model misspecification, mitigated
  structurally by the blast-radius cap, not statistically.
\item \textbf{The advantage condition is frequency-free but budget-dependent.} It uses
  only an upper bound $\bar\varepsilon$ on fatal mass, never a point estimate ---
  the minimax robustness we wanted. But it still trusts that budget, and it
  assumes i.i.d.\ periods; clustered fatal events (regimes, cascades), where
  frequency estimates fail worst, are unmodeled.
\item \textbf{The adversary moves once.} The converse is a one-shot game. The iterated
  game against a learning detector --- and whether detector \emph{diversity} helps the
  way portfolio diversity does --- is open.
\item \textbf{The price of determinism is exact but not computable-by-us.} The
  deterministic many-direction value is a max-min over a non-convex joint ROC
  set; we resolve it for randomized play (the duality of \S\ref{sec:taxonomy}) but have neither
  a simplification nor a hardness proof for the deterministic case --- the
  nearby classifier-rejector hardness results suggest NP-hardness, and we
  have not shown it.
\item \textbf{``Removable'' may be one concept or three.} Permanent removal (buy the
  capability --- the defect \emph{becomes} a capability), per-event removal (escalate
  this once), and interpretive removal (the sealed-frame case, where the seal
  is never lifted and removal means compensating its one fatal reading) share a
  single manifest field in \S\ref{sec:agentworld} but may deserve separate treatment. The
  observation/capacity split (\S\ref{sec:taxonomy}) is first evidence that the taxonomy is real;
  the removability spectrum is a second, orthogonal axis, and we have unified
  neither.
\item \textbf{The human case is deliberately excluded.} The model prices artifacts.
  Applying ``a person's deficiency is an advantage'' to people imports ethical
  questions --- who owns the compensation channel, autonomy versus dependency ---
  that no theorem here resolves, and we do not pretend otherwise.
\end{itemize}

\section{Conclusion}\label{sec:conclusion}

A deliberately narrow system with a tripwire and an escalation path is a
common engineering pattern, usually justified by intuition. This paper gives
it an economics and a limit. The economics: keeping a deficiency is a priced
position, advantageous when specialization gain outruns the insured tail. The
limit: removal is possible exactly when the detector sits outside the defect,
and impossible --- regardless of switch quality --- when the defect and its harm
share a substrate, because then the rate the feature helps and the rate it
harms are the same number. Between the two sits an exact accounting: one
value formula prices every detector class, the shortfall against the ideal
decomposes into two named obstructions with a coin buying back one and never
the other, and the whole architecture can be certified from data at a
training bill linear in the severity it insures. The single design rule that
follows is sharper than
``add a monitor'': \emph{narrow the action all you like, but never let the narrowing
reach the detector.} Perception must stay open even where action is closed.

% ===========================================================================
\appendix

\section*{Overview of the appendix}

This appendix proves both directions of the detector-placement iff
(Conjectures B and A of the project's formalization notes), after an honest
repair that the proof itself forced. The proofs are elementary once the model
is right; per the project's convention, we say so --- the content is in the
definitions and in the quantifier structure, not in analytic difficulty.

\textbf{What is proved here:}
\begin{itemize}
\item Lemma 1 (coupling): inside an observation defect, the detector's gain-capture
  rate and its fatal-miss rate are the same number.
\item Theorem 1 (observation defects): the specialization premium collapses to
  zero, \emph{unconditionally} --- transductive access does not help.
\item Theorem 1$'$ (value formula): the exact within-defect value, with the
  removability coefficient $\sigma_0$ (residual separability) governing the collapse.
\item Theorem 2 (capacity defects): the premium collapses in the inductive regime,
  and the inductive condition is exactly a minimax gap --- transduction closes
  it, matching \citet{goldwasser2020}.
\item \textbf{Theorem 2$'$ (capacity value formula, Appendix~\ref{app:thm2prime}):} for \emph{any} detector class, the
  premium is the support function of the class's ROC set at an economic price
  vector, with removability coefficient the zero-leak capture capacity
  $\bar\sigma_0(H)$. This subsumes Theorem 1$'$ --- observation and capacity
  defects obey \textbf{one value formula} --- and relocates the split entirely to
  joint realizability across confusable directions (the transduction gap).
\item \textbf{The deficit characterized (Appendix~\ref{app:deficit}):} the joint-realizability deficit is the
  limiting transduction gap in premium units; it decomposes into exactly two
  obstructions (cross-leak, non-closure); exact formulas for monotone classes
  (``global paranoia'') and halfspaces (convex non-separability); and Theorem R
  bounds it by a direction router's confusion mass at the same two prices ---
  the router is itself a detector, the question recurses, and the recursion
  is the observation/capacity split in its third formulation.
\item \textbf{Theorem M (mixed obstructions, Appendix~\ref{app:thmM}):} the $k$-direction game, with per-task
  randomization, collapses by LP duality to a \emph{single} Theorem 2$'$ instance at
  a least-favorable mixture of directions; the transduction gap splits
  additively into irreducible cross-leak plus the closure deficit --- a coin
  buys back the gluing failure, never the cross-leak. XOR witnesses both
  halves at exactly $A/2$ each.
\item Corollary 3 (growth form): under multiplicative dynamics, survival and
  specialization gain cannot coexist inside the defect.
\item \textbf{Theorem A (achievability, Appendix~\ref{app:thmA}):} a detector \emph{outside} the defect earns a
  worst-case premium bounded below by an explicit $\Pi^* > 0$ under the
  advantage condition; the premium identity has three per-unit-priced terms.
  A meets B exactly at the $L \to \infty$ boundary, and the two give the
  \textbf{full iff} (Corollary A+B).
\item \textbf{Theorem E/E$'$ (end-to-end, Appendix~\ref{app:thmE}):} the detector learned from stratified
  samples of a declared cell cover, with certified rates. The rare-event form
  certifies the fatal side at a one-time training bill \emph{linear} in severity
  $L$ (the agnostic route costs $L^2$); amortized, learning preserves the
  advantage condition. Default-deny gating makes the certificate fail-safe
  against cover incompleteness; the residual risk is mis-declaration.
\end{itemize}

\textbf{What the proof forced us to admit:} the drafted definition of ``removable''
was vacuous, and the drafted Conjecture B was false under it (Appendix~\ref{app:vacuous}). The
repaired notion is \emph{profitable removability}.

\textbf{Still open (Appendix~\ref{app:open}):} the iterated adversary, the deterministic max-min's
complexity, iterated routing, and beyond-mixture uncertainty classes.

\section{Model}\label{app:model}

Task space $(X, \Sigma)$, one task per period. Competence set $C \in \Sigma$;
fatal exposure $E := F \setminus C \in \Sigma$, disjoint from $C$. Per-task
payoffs for the two actions:

\begin{center}
\begin{tabular}{lcc}
\toprule
 & $x \in C$ & $x \in E$ \\
\midrule
\textbf{act} & $+g$ & $-L$ \\
\textbf{escalate} & $-p$ & $-p$ \\
\bottomrule
\end{tabular}
\end{center}

\noindent
with $g, p > 0$ and $L$ the fatal loss; \textbf{standing assumption:} $L > p$
(a missed fatal costs more than an escalation --- otherwise always-act
dominates and there is nothing to insure). Tasks outside $C \cup E$ form the
\textbf{benign region} $N := X \setminus (C \cup E)$: acting there pays $-\ell$
with $0 \leq \ell \leq p$, escalating pays $-p$; benign mass enters no result
below except through Theorem A's accounting. A detector is a measurable
$d : X \to \{a, e\}$; the agent's per-period expected payoff under task
distribution $P$ is $G_P(d) = \mathbb{E}_P[\text{payoff}]$. The detector rent
$c_d$ is a constant shift within any fixed detector class and is dropped. A
third action \emph{decline} (payoff $0$, task not done) changes nothing below
except replacing the forfeit value $-p$ by $0$; we note where.

Throughout, fix $\bar\varepsilon \in (0,\, g/(g+p))$ --- the fatal-mass budget ---
and define the \textbf{critical loss}
$$L^* \;:=\; \frac{(1-\bar\varepsilon)\,g + p}{\bar\varepsilon}.$$

\textbf{Two formalizations of the defect.}
\begin{itemize}
\item \textbf{Observation defect.} The agent perceives tasks only through a measurable
  map $\varphi : X \to Z$; its detector must be $\sigma(\varphi)$-measurable.
  Write $\mathcal{D}_R$ for this class. (Coarsened observation channel;
  covers pruned/quantized perception.)
\item \textbf{Capacity defect.} The agent perceives $x$ fully but its detector must
  come from a restricted class $H \subseteq \{a,e\}^X$ containing the two
  constant maps. (Restricted function class.)
\end{itemize}

\textbf{Adversarial families.} For probability measures $\nu_0$ (supported in $C$)
and $\nu_1$ (supported in $E$), define the mixture family
$$\mathcal{F}(\nu_0, \nu_1) \;:=\; \{\, P_\lambda = (1-\lambda)\nu_0 + \lambda \nu_1 \;:\; \lambda \in [0, \bar\varepsilon] \,\}.$$

All results are stated for uncertainty classes $\mathcal{P}$ containing such a
family; protection claims are always relative to the declared class, never to
nature (the formalization notes' honesty rule).

\section{The definition was vacuous --- profitable removability}\label{app:vacuous}

The project's formalization notes defined a defect \emph{removable} if some policy keeps $G$
bounded below uniformly over $\mathcal{P}$. Writing the proof exposed the
flaw: \textbf{the constant detector $d \equiv e$ (always escalate) lies in every
detector class} --- constants are measurable for any $\sigma(\varphi)$ and were
assumed in $H$ --- \textbf{and bounds $G \geq -p$ regardless of the defect.} Under
the drafted definition every defect is removable and Conjecture B was
vacuously false, independently of the Goldwasser escape already recorded in
the project's literature notes. Survival was never the question; \emph{paid-for}
survival is.

\begin{ndef}{Definition}[specialization premium]
For a detector class $\mathcal{D}$ and family $\mathcal{P}$,
$$\Pi(\mathcal{D}) \;:=\; \sup_{d \in \mathcal{D}} \inf_{P \in \mathcal{P}} G_P(d) \;-\; \inf_{P \in \mathcal{P}} G_P(d \equiv e),$$
the worst-case value above the always-escalate baseline. (With \emph{decline}
available, the baseline is $0$; all premium statements are unchanged.)
\end{ndef}

\begin{ndef}{Definition}[profitably removable]
The defect is \emph{profitably removable}
on $\mathcal{P}$ if $\Pi(\mathcal{D}_{\text{defect}}) > 0$: some within-defect
policy retains part of the specialization gain while avoiding ruin.
\end{ndef}

The theorems below say: with a confounded detector the premium is zero (or
$\to 0$), while an out-of-defect detector on the same family earns premium
$(1-\bar\varepsilon)(g+p)$. That is the honest, non-vacuous form of ``the
defect must not include the detector.''

\section{Lemma 1 --- the coupling lemma}\label{app:coupling}

\begin{nthm}{Lemma 1}
Let $d \in \mathcal{D}_R$ (observation defect) and let
$\nu_0, \nu_1$ satisfy $\varphi_* \nu_0 = \varphi_* \nu_1$ (equal
pushforwards: the coarsened view cannot distinguish the safe ensemble from
the fatal one). Then
$$\underbrace{\nu_0(d = a)}_{\text{gain-capture rate } q_0} \;=\; \underbrace{\nu_1(d = a)}_{\text{fatal-miss rate } q_1}.$$
\end{nthm}

\begin{proof}
By Doob--Dynkin, $\sigma(\varphi)$-measurability of $d$ gives a
measurable $d' : Z \to \{a,e\}$ with $d = d' \circ \varphi$. Then
$\nu_i(d = a) = \nu_i(\varphi^{-1}(d'{=}a)) = (\varphi_*\nu_i)(d'{=}a)$, and
the two pushforwards are equal.
\end{proof}

\textbf{Reading.} Inside the defect, \textbf{the miss rate \emph{is} the capture rate}: one
scalar $q := q_0 = q_1$ prices both. Every unit of specialization gain the
detector lets through carries a proportional unit of fatal exposure --- the
two cannot be decoupled by any cleverness downstream of $\varphi$, because
they are literally the same number. Outside the defect they decouple
completely ($q_0 = 1$, $q_1 = 0$ is available). This is the formal content of
``shared difficulty function $\Rightarrow$ correlated failure'' \citep{eckhardt1985,littlewood2012}, upgraded from a correlation model to an identity.

\textbf{Quantitative version.} Without equal pushforwards, let
$\tau := \mathrm{TV}(\varphi_*\nu_0, \varphi_*\nu_1)$. Then for every
$d \in \mathcal{D}_R$: $\;q_0 - q_1 \leq \tau$ (immediate from the definition
of total variation). The retained distinguishability $\tau$ is the \emph{only}
wedge the restricted detector can drive between capture and miss.

\section{Theorem 1 --- observation defects: unconditional collapse}\label{app:thm1}

\begin{nthm}{Theorem 1}
Suppose $\varphi_*\nu_0 = \varphi_*\nu_1$ and
$\mathcal{P} \supseteq \mathcal{F}(\nu_0, \nu_1)$. Then for every
$L \geq L^*$:
\begin{enumerate}
\item \textbf{(Collapse.)} $\displaystyle \sup_{d \in \mathcal{D}_R} \inf_{\lambda} G_{P_\lambda}(d) = -p$; for $L > L^*$ it is attained only by policies with $q = 0$ (total forfeit on the family; at $L = L^*$ every $q$ attains it). Hence $\Pi(\mathcal{D}_R) = 0$.
\item \textbf{(Transduction is powerless.)} Minimax equality holds:
   $\displaystyle \inf_{\lambda} \sup_{d \in \mathcal{D}_R} G_{P_\lambda}(d) = -p$ as well. Even a detector chosen \emph{after} seeing the deployment distribution earns zero premium.
\item \textbf{(Separation.)} The unrestricted detector $d^*= \mathbf{1}[x \in C]\cdot a$ achieves $\inf_\lambda G_{P_\lambda}(d^*) = (1-\bar\varepsilon)g - \bar\varepsilon p > 0$, i.e. premium $(1-\bar\varepsilon)(g+p)$. This is exactly optimal: $\sup_{d \text{ } \Sigma\text{-meas.}} \inf_\lambda G_{P_\lambda}(d) = (1-\bar\varepsilon)g - \bar\varepsilon p$.
\end{enumerate}
\end{nthm}

\begin{proof}
Let $d \in \mathcal{D}_R$ with common rate $q$ (Lemma 1). Direct
computation:
$$G_{P_\lambda}(d) \;=\; (1-\lambda)\big[qg - (1-q)p\big] + \lambda\big[-qL - (1-q)p\big] \;=\; q\big[(1-\lambda)g - \lambda L\big] - (1-q)\,p.$$

(1) The expression is linear in $q$ and decreasing in $\lambda$ for $q>0$, so
$\inf_\lambda$ sits at $\lambda = \bar\varepsilon$, giving
$q\big[(1-\bar\varepsilon)g - \bar\varepsilon L\big] - (1-q)p$. For
$L \geq L^*$ the bracket is $\leq -p$, so the whole expression is maximized
at $q = 0$ with value $-p$; for $L > L^*$ the bracket is $< -p$ and any
$q > 0$ does strictly worse (and $\to -\infty$ as $L \to \infty$), while at
$L = L^*$ the bracket equals $-p$ and every $q$ ties.

(2) For fixed $\lambda$, $\sup_{d}$ over $\mathcal{D}_R$ is the same linear
program in $q \in [0,1]$ (both endpoints are realized by the constant
detectors, which lie in $\mathcal{D}_R$):
$\sup_d G_{P_\lambda} = \max\big(-p,\; (1-\lambda)g - \lambda L\big)$. This is
decreasing in $\lambda$; the adversary plays $\lambda = \bar\varepsilon$ and
for $L \geq L^*$ the max is $-p$.

(3) For arbitrary $\Sigma$-measurable $d$ let $q_0 := \nu_0(d{=}a)$,
$q_1 := \nu_1(d{=}a)$ --- now free to differ. $G_{P_\lambda}(d) = (1-\lambda)[q_0 g - (1-q_0)p] + \lambda[-q_1 L - (1-q_1)p]$
is maximized by $q_1 = 0$, $q_0 = 1$, i.e. by $d^*$, giving
$(1-\lambda)g - \lambda p$, whose infimum over the family is at
$\lambda = \bar\varepsilon$.
\end{proof}

\textbf{Remarks.}
\begin{itemize}
\item Point (2) is stronger than the repaired Conjecture B asked for: for
  observation defects the inductive-regime condition is \emph{unnecessary}.
  Transduction tells you \emph{that} fatal mass is present ($\lambda$), never
  \emph{which} task is fatal --- the information was destroyed per-task by
  $\varphi$, before any distributional knowledge can act. Knowing $\lambda$,
  the optimal within-defect play is precisely a Chow rule on the coarsened
  posterior --- and at the confounded points that posterior equals $\lambda$
  itself, forcing all-or-nothing behavior: acting everywhere or escalating
  everywhere. Chow's rule executes the collapse; it cannot prevent it.
\item The gap between inside and outside is $(1-\bar\varepsilon)(g+p)$ --- the
  entire specialization premium, independent of $L$ beyond the threshold
  $L^*$.
\end{itemize}

\section{Theorem 1\prm{} --- exact value and the removability coefficient}\label{app:thm1prime}

Drop the equal-pushforward assumption. Write $m_i := \varphi_* \nu_i$ and
define the \textbf{residual separability}
$$\sigma_0 \;:=\; \sup\{\, m_0(S) \;:\; S \subseteq Z \text{ measurable},\; m_1(S) = 0 \,\}$$
--- the share of the safe ensemble recoverable at \emph{zero} coarsened fatal risk.

\begin{nthm}{Theorem 1$'$}
On the family $\mathcal{F}(\nu_0,\nu_1)$, within-defect
detectors correspond exactly to measurable act-regions $S \subseteq Z$, and
$$\sup_{d \in \mathcal{D}_R} \inf_{\lambda} G_{P_\lambda}(d) \;=\; \sup_{S \subseteq Z}\; \Big\{ (1-\bar\varepsilon)\big[m_0(S)\,g - (1{-}m_0(S))\,p\big] \;-\; \bar\varepsilon\big[m_1(S)\,L + (1{-}m_1(S))\,p\big] \Big\},$$
a Neyman--Pearson problem on the coarsened likelihood ratio $dm_1/dm_0$.
Consequently, as $L \to \infty$,
$$\Pi(\mathcal{D}_R) \;\longrightarrow\; \sigma_0 \cdot (1-\bar\varepsilon)(g+p) \;=\; \sigma_0 \cdot \Pi(\Sigma\text{-measurable}).$$
\end{nthm}

\begin{proof}
The correspondence $d \leftrightarrow S = \{d' = a\}$ is Lemma 1's
factorization; the displayed value is the direct computation of
$\inf_\lambda G$ (infimum again at $\lambda = \bar\varepsilon$ for any $S$
with $m_1(S) \geq 0$, since the $\lambda$-coefficient is negative whenever
$S$ has any mass). For $L \to \infty$, any $S$ with $m_1(S) > 0$ has value
$\to -\infty$, so the supremum concentrates on $\{S : m_1(S) = 0\}$, over
which the value is increasing in $m_0(S)$ with supremum
$(1-\bar\varepsilon)[\sigma_0 g - (1-\sigma_0)p] - \bar\varepsilon p = -p + \sigma_0(1-\bar\varepsilon)(g+p)$.
\end{proof}

\textbf{Reading.} $\sigma_0$ is the fraction of the specialization premium that
survives the coarsening: $\sigma_0 = 1$ recovers the full premium,
$\sigma_0 = 0$ is Theorem 1's collapse. \textbf{For observation defects on mixture
families this settles the ``iff'' direction of the slogan}: the defect is
profitably removable in the large-$L$ limit iff the restriction retains a
region where safe mass lives at zero fatal mass --- iff the detector-relevant
distinction survives outside the defect. Removability is not a yes/no after
all (research backlog \#6): it is the coefficient $\sigma_0$.

\section{Theorem 2 --- capacity defects: collapse in the inductive regime, and the minimax gap}\label{app:thm2}

Now the detector sees $x$ fully but is confined to
$H \subseteq \{a,e\}^X$ (constants included). Fix distinct pairs
$(x_0^i, x_1^i)_{i=1}^k$ with $x_0^i \in C$, $x_1^i \in E$, and let
$\mathcal{P} = \bigcup_i \mathcal{F}(\delta_{x_0^i}, \delta_{x_1^i})$.

\begin{itemize}
\item \textbf{Confounding condition:} every $h \in H$ pools some pair:
  $\forall h\, \exists i : h(x_0^i) = h(x_1^i)$.
\item \textbf{Resolution condition:} every pair is resolved by some member:
  $\forall i\, \exists h_i \in H : h_i(x_0^i) = a,\; h_i(x_1^i) = e$.
\end{itemize}

\begin{nthm}{Theorem 2}
Under the confounding condition, for $L \geq L^*$:
\begin{enumerate}
\item \textbf{(Inductive collapse.)} $\displaystyle \sup_{h \in H} \inf_{P \in \mathcal{P}} G_P(h) = -p$; hence $\Pi(H) = 0$ in the inductive regime.
\item \textbf{(Transductive escape.)} Under the resolution condition additionally,
   $\displaystyle \inf_{P \in \mathcal{P}} \sup_{h \in H} G_P(h) \geq (1-\bar\varepsilon)g - \bar\varepsilon p$, so the minimax gap is at least $(1-\bar\varepsilon)(g+p)$ --- the entire premium.
\end{enumerate}
\end{nthm}

\begin{proof}
(1) Fix $h$; pick a pooled pair $i = i(h)$. On
$P^i_{\bar\varepsilon}$: if $h$ acts on both members,
$G = (1-\bar\varepsilon)g - \bar\varepsilon L \leq -p$ for $L \geq L^*$; if
it escalates both, $G = -p$. Either way
$\inf_P G_P(h) \leq -p$, and the constant $e$ attains $-p$. (2) For any
$P^i_\lambda$, play the resolving $h_i$:
$G = (1-\lambda)g - \lambda p \geq (1-\bar\varepsilon)g - \bar\varepsilon p$.
\end{proof}

\textbf{Remarks.}
\begin{itemize}
\item \textbf{The inductive condition of the repaired Conjecture B is now identified
  formally: it is the order of quantifiers.} Inductive = detector committed
  before the deployment distribution ($\sup_h \inf_P$); transductive =
  detector chosen with knowledge of it ($\inf_P \sup_h$). For capacity
  defects the gap between the two is the whole premium --- this is the toy form
  of exactly the mechanism by which \citet{goldwasser2020}
  escape: unlabeled deployment data moves the system from the first value to
  (toward) the second. For observation defects, Theorem 1(2) says the gap is
  zero. \textbf{Backlog \#6's ``one concept or three'' receives its first hard
  evidence: observation and capacity defects are provably different along the
  transduction axis.}
\item Realizability of the conditions is generic, not exotic. Example: $H$ =
  threshold detectors on a one-dimensional confidence score $s$, with
  safe/fatal points interleaved in $s$ (scores $1, 2, 3, 4$ for
  $x_0^1, x_1^1, x_0^2, x_1^2$): every threshold pools one of the two pairs,
  yet each pair is resolved by some threshold. Interleaved confidence is the
  norm for restricted scores --- this is Ulmer--Cin\`a's ReLU-confidence failure
  \citep{ulmer2021} as a two-pair combinatorial fact.
\end{itemize}

\section{Theorem 2\prm{} --- the capacity value formula, and the unification}\label{app:thm2prime}

Theorem 2 gave the capacity-defect collapse as an inequality under a
combinatorial condition. Here is the exact value, for an arbitrary class --- and
it turns out to be the \emph{same formula} as Theorem 1$'$, which relocates the
observation/capacity split.

\textbf{Setup.} Detector class $H \subseteq \{a,e\}^X$ (constants included),
ensemble pair $\nu_0$ (supported in $C$), $\nu_1$ (supported in $E$), family
$\mathcal{F}(\nu_0,\nu_1)$. Each $h \in H$ has an \textbf{operating point}
$\big(q_0(h), q_1(h)\big) := \big(\nu_0(h{=}a),\, \nu_1(h{=}a)\big)$ ---
capture rate and leak rate; note $q_1$ \emph{is} the miss rate $\mu$ on the fatal
ensemble and $1 - q_0$ the false-alarm rate on the safe one. Call
$\mathrm{ROC}(H) := \{(q_0(h), q_1(h)) : h \in H\} \subseteq [0,1]^2$ the
class's \textbf{ROC set} against the pair. Write the \textbf{price vector}
$$A := (1-\bar\varepsilon)(g+p) \quad (\text{per unit captured}), \qquad B := \bar\varepsilon\,(L-p) \quad (\text{per unit leaked}).$$

\begin{nthm}{Lemma 2}[operating-point reduction]
For every $h \in H$, the worst-case
premium over the family depends on $h$ only through its operating point:
$$\Pi(h) \;=\; A\,q_0(h) \;-\; B\,q_1(h).$$
\end{nthm}

\begin{proof}
$G_{P_\lambda}(h) = (1-\lambda)[q_0 g - (1-q_0)p] + \lambda[-q_1 L -
(1-q_1)p]$, which is nonincreasing in $\lambda$ (strictly, unless
$q_0 = q_1 = 0$), so the infimum sits at $\lambda = \bar\varepsilon$;
subtracting the baseline $-p$ and collecting terms gives the display.
\end{proof}

\begin{nthm}{Theorem 2$'$}[capacity value formula]
For any $H$ containing the constants,
$$\Pi(H) \;=\; \sup_{(q_0, q_1) \in \mathrm{ROC}(H)} \big[ A\,q_0 - B\,q_1 \big]$$
--- the \textbf{support function of the class's ROC set at the price vector
$(A, -B)$}. Consequently:
\begin{enumerate}
\item The supremum is approached on the Pareto frontier of $\mathrm{ROC}(H)$ ---
   the class-restricted ROC curve --- and the within-class optimal detector is
   the class-restricted Neyman--Pearson point at slope $B/A$: Chow's rule
   again, executed inside the class.
\item As $L \to \infty$, $\;\Pi(H) \to A \cdot \bar\sigma_0(H)$, where
   $$\bar\sigma_0(H) \;:=\; \lim_{t \downarrow 0}\; \sup\{\, q_0(h) : h \in H,\; q_1(h) \leq t \,\}$$
   is the \textbf{zero-leak capture capacity} of the class (the limit exists by
   monotonicity in $t$).
\end{enumerate}
\end{nthm}

\begin{proof}
The formula is Lemma 2 plus taking the supremum; $(0,0) \in
\mathrm{ROC}(H)$ via the constant $e$, so $\Pi(H) \geq 0$ automatically. For
(2): upper bound --- for any $h$, $\Pi(h) \leq \min\big(A\,q_0(h),\; A -
B\,q_1(h)\big)$, so any $h$ competing with the limit must have $q_1(h) \leq
A/B \to 0$ and hence $q_0(h) \leq \sup\{q_0 : q_1 \leq A/B\} \to
\bar\sigma_0(H)$. Lower bound --- couple $t$ to $L$: since
$\sup\{q_0(h) : q_1(h) \leq t\} \geq \bar\sigma_0(H)$ for every $t > 0$ (the
sup is nonincreasing as $t \downarrow 0$), take $t(L) := L^{-2}$ and pick
$h_L$ with $q_1(h_L) \leq t(L)$ and $q_0(h_L) \geq \bar\sigma_0(H) - \eta$;
then $\Pi(H) \geq \Pi(h_L) \geq A\,(\bar\sigma_0(H) - \eta) - B\,t(L)$ with
$B\,t(L) = \bar\varepsilon(L-p)\,L^{-2} \to 0$.
\end{proof}

\textbf{Remark (closure).} $\bar\sigma_0(H) \geq \sigma_0(H) := \sup\{q_0 : q_1 =
0\}$, with equality when the ROC set is closed --- finite classes, compact
operating sets --- and possibly strict otherwise (a class whose every
non-constant member leaks a little, with no zero-leak limit point, has
$\sigma_0 = 0 < \bar\sigma_0$). For $\sigma$-algebra classes the two coincide: from
$S_n$ with $m_1(S_n) \to 0$ summably and $m_0(S_n) \geq c$, the set $S =
\limsup S_n$ has $m_1(S) = 0$ and $m_0(S) \geq c$ by reverse Fatou. So
Theorem 1$'$'s $\sigma_0$ needed no correction.

\begin{nthm}{Corollary}[unification]
Theorem 1$'$ is the special case $H = $ the
$\sigma(\varphi)$-measurable detectors: by Lemma 1's factorization,
$\mathrm{ROC}(H_\varphi) = \{(m_0(S), m_1(S)) : S \subseteq Z\}$, the support
function reproduces Theorem 1$'$'s value, and $\bar\sigma_0 = \sigma_0$ by the
closure remark. \textbf{Observation and capacity defects obey one value formula:
premium $=$ support function of the detector class's ROC set at the economic
price vector; removability coefficient $=$ zero-leak capture capacity.} What
distinguishes the two defect types is only how the ROC set arises --- the image
of a coarsened $\sigma$-algebra versus the trace of a function class.
\end{nthm}

\textbf{Where the split actually lives.} On a \emph{single} mixture family the
transduction gap vanishes for \textbf{every} class: each $h$'s infimum over
$\lambda$ sits at $\bar\varepsilon$, so
$\inf_\lambda \sup_h G \leq \sup_h G_{P_{\bar\varepsilon}} = \sup_h
\inf_\lambda G$, and the minimax inequality closes the other side. The gap of
Theorem 2 is therefore a \textbf{multi-direction phenomenon}: with a union of
confusable families the inductive value is the \emph{joint} max-min
$\sup_h \min_j [A\,q_0^j(h) - B\,q_1^j(h)]$, the transductive value is the
per-direction $\min_j \sup_h$, and the gap measures \textbf{joint-realizability
deficit} --- whether one member of the class can serve all confusable
directions at once. Theorem 2's confounding/resolution conditions are the
extreme case (each direction resolvable, no member resolves all). This
sharpens the taxonomy result: the observation/capacity split is \emph{not} about
how a detector is priced --- the formula above is common --- but about whether
deployment knowledge lets the class pick its member per direction. The
limiting coefficient for a union family is accordingly
$\bar\sigma_0(H; \{\nu^j\}) = \lim_{t \downarrow 0} \sup\{\min_j q_0^j(h) :
h \in H,\; \max_j q_1^j(h) \leq t\}$, which recovers Theorem 2(1)'s collapse
as $\bar\sigma_0 = 0$ under confounding.

\section{The joint-realizability deficit, characterized}\label{app:deficit}

Theorem 2$'$ located the transduction gap in joint realizability. This section
makes that quantitative: what the deficit \emph{is}, which two obstructions
produce it, exact formulas for the natural classes, and a general bound that
turns out to be the theory applied to itself.

\textbf{Setup.} Directions $j = 1, \dots, k$: ensemble pairs $(\nu_0^j, \nu_1^j)$,
union family $\mathcal{P} = \bigcup_j \mathcal{F}(\nu_0^j, \nu_1^j)$. Write
$\bar\sigma^j(H)$ for the per-direction zero-leak capture capacity and
$\bar\sigma^{\mathrm{jt}}(H) := \lim_{t \downarrow 0} \sup\{\min_j q_0^j(h) :
h \in H,\ \max_j q_1^j(h) \leq t\}$ for the joint one (Appendix~\ref{app:thm2prime}). Write
$\Pi^{\mathrm{ind}}_L(H) := \sup_{h \in H} \min_j \big[A\,q_0^j(h) - B\,q_1^j(h)\big]$
for the \textbf{inductive} premium (one detector must serve every direction) and
$\Pi^{\mathrm{tr}}_L(H) := \min_j \sup_{h \in H} \big[A\,q_0^j(h) - B\,q_1^j(h)\big]$
for the \textbf{transductive} one (the detector may be chosen after the
direction is revealed), as in Appendix~\ref{app:thm2prime}. Define the
\textbf{joint-realizability deficit}
$$\Delta(H) \;:=\; \min_j \bar\sigma^j(H) \;-\; \bar\sigma^{\mathrm{jt}}(H) \;\in\; [0, 1],$$
(nonnegative since any jointly feasible $h$ is feasible per direction). By
Theorem 2$'$'s limits applied per-direction and jointly, the transduction gap
satisfies $\Gamma_L := \Pi^{\mathrm{tr}}_L - \Pi^{\mathrm{ind}}_L \to
A \cdot \Delta(H)$ as $L \to \infty$: \textbf{the deficit is the limiting gap in
premium units.}

\begin{nthm}{Proposition 4}[the two obstructions]
If for each $j$ and every $t > 0$
there is $h_j \in H$ with $q_0^j(h_j) \geq \bar\sigma^j - \eta$ and
$q_1^i(h_j) \leq t$ for \textbf{all} $i$ (\emph{cross-safe} local solutions), and $H$
contains a member whose act region \emph{is} the union of the $h_j$'s act
regions (\emph{union closure} over them), then $\Delta(H) = 0$.
\end{nthm}

\begin{proof}
The union member captures $q_0^j \geq \bar\sigma^j - \eta$ on each
$j$ and leaks at most $kt$ on each $i$; let $t, \eta \downarrow 0$.
\end{proof}

So every positive deficit is attributable to \textbf{cross-leak} (direction $i$'s
solution acts on direction $j$'s fatal mass) or \textbf{non-closure} (the class
cannot glue its local solutions), and the two are separated by the examples
below: thresholds fail only by cross-leak, halfspaces only by non-closure.

\begin{nthm}{Proposition 5}[monotone classes --- exact formula, ``global paranoia'']
Let $H = \{ \text{act iff } s(x) \geq \theta \}$ for a score $s$ (act regions
nested, hence union-closed). Let $T_j := \inf\{\theta : \nu_1^j(s \geq
\theta) = 0\}$ be direction $j$'s fatal ceiling and $T := \max_j T_j$ the
global one. Then
$$\bar\sigma^j = \nu_0^j(s > T_j), \qquad \bar\sigma^{\mathrm{jt}} = \min_j\, \nu_0^j(s > T), \qquad \Delta = \min_j \nu_0^j(s > T_j) - \min_j \nu_0^j(s > T)$$
(with $>$ relaxed to $\geq$ at ceilings carrying no fatal atom). The deficit
is the safe mass trapped \textbf{between each direction's own fatal ceiling and the
global one}: a single threshold must be paranoid enough for the worst
direction and pays that paranoia in every direction. \emph{Point-pair corollary:}
$\Delta \in \{0,1\}$, and $\Delta = 1$ iff the pairs are correctly ordered
but \textbf{interleaved} --- every $s(x_0^j) > s(x_1^j)$, yet some safe point scores
below another direction's fatal point. (Theorem 2's threshold example, now
exact; Ulmer--Cin\`a's interleaved-confidence failure is this formula.)
\end{nthm}

\begin{proof}
Nested act regions make every supremum one-dimensional in $\theta$;
zero leak on $j$ forces $\theta > T_j$ (all $i$: $\theta > T$), and
$\{s \geq \theta\} \downarrow \{s > T\}$ as $\theta \downarrow T$ gives the
right-limits.
\end{proof}

\begin{nthm}{Proposition 6}[halfspaces --- the deficit is non-separability]
Let $H$ =
halfspace act regions in $\mathbb{R}^n$, directions given by point pairs.
Each pair is \emph{always} individually resolvable (two points), so
$\min_j \bar\sigma^j = 1$; and $\bar\sigma^{\mathrm{jt}} = 1$ iff the safe
points $\{x_0^j\}$ are strictly linearly separable from the fatal points
$\{x_1^j\}$, else $0$. Hence $\Delta \in \{0,1\}$ with
$$\Delta = 1 \iff \mathrm{conv}\{x_0^j\} \cap \mathrm{conv}\{x_1^j\} \neq \emptyset.$$
The minimal witness is the XOR configuration ($k = 2$, $n \geq 2$).
Halfspaces are cross-safe here (a pair's separator can be chosen missing the
other points) but not union-closed --- the complementary failure mode to
Proposition 5. \qquad $\square$
\end{nthm}

\textbf{Remark (no single capacity number).} If $H$ shatters the $2k$ direction
points, $\Delta = 0$ trivially --- so a positive deficit certifies that the
directions witness a non-shattered set. But the converse fails badly:
thresholds (VC 1), intervals (VC 2), and halfspaces (VC $n{+}1$) all admit
$\Delta = 1$ at $k = 2$ under adversarial placement. The deficit is a
function of the \textbf{geometry of the directions relative to the class}, not of
class capacity alone; shattering gives sufficiency, never a characterization.

\begin{nthm}{Theorem R}[router bound --- the theory applied to itself]
Let
$r : X \to [k]$ be any \emph{direction router} with confusion mass
$\rho := \max_j \max\big(\nu_0^j(r \neq j),\ \nu_1^j(r \neq j)\big)$, and let
the gated composite $h_r(x) := h_{r(x)}(x)$ be admissible (i.e. play in the
augmented class $R \ltimes H^k$). Then at \textbf{every} $L$:
$$\Gamma_L(R \ltimes H^k) \;\leq\; \rho \cdot (A + B).$$
\end{nthm}

\begin{proof}
Pick per-direction detectors $h_j$ within $\eta$ of the transductive
optimum at the current $L$. On direction $j$ the composite's capture drops by
at most $\nu_0^j(r \neq j) \leq \rho$ and its leak rises by at most
$\nu_1^j(r \neq j) \leq \rho$, so its premium is within
$\rho(A + B) + \eta$ of $\Pi^{\mathrm{tr}}_L$.
\end{proof}

Three consequences:

\begin{enumerate}
\item \textbf{The deficit is priced by the same two prices} --- router confusion on
   safe mass costs $A$ per unit, on fatal mass $B$ per unit. Since
   $B = \bar\varepsilon(L - p)$ grows with $L$, a useful bound at large $L$
   needs router confusion $O(1/L)$ on the fatal ensembles --- \emph{exactly}
   Theorem A's law for the primary detector, at the same
   $O(\log L)$ hypothesis-testing rent. The router \textbf{is} a detector --- of
   directions instead of fatality --- and obeys detector economics.
\item \textbf{The escape-route question recurses.} ``Can the class serve all
   directions at once?'' reduces to ``can the direction be detected per task?''
   --- a second-order instance of the original problem. The recursion
   terminates in one of two ways, and this is the observation/capacity split
   in its third and sharpest formulation: for a \textbf{capacity defect} the
   recursion bottoms out in a cheap router (directions distinguishable from
   the task; deficit $\leq \rho(A{+}B)$, bought at log-rent); for an
   \textbf{observation defect} the router would need exactly the information the
   coarsening destroyed --- its confusion is bounded \emph{below} by the coupling
   lemma --- and the recursion halts at level one with nothing to buy.
\item \textbf{Fleet corollary} (research backlog \#5): a menu of $k$ specialists plus a
   router is one generalist detector, and $\rho(A + B)$ is precisely what the
   fleet loses to the ideal; the composition question inherits the whole
   pricing apparatus.
\end{enumerate}

The bound is stated for the augmented class: it characterizes what
\emph{composition closure buys}, while Propositions 5--6 give the exact deficit of
the raw classes. The mixed case --- cross-leak and non-closure together, on
ensemble directions --- is Appendix~\ref{app:thmM}.

\section{Theorem M --- the mixed-obstruction deficit: duality and the price of determinism}\label{app:thmM}

Where Appendix~\ref{app:deficit} characterized the deficit with one obstruction acting alone, this
section gives the general case. The conjectured shape (a Neyman--Pearson
problem over the joint ROC set) is correct, and its \emph{dual} is the theorem:
the $k$-direction game collapses to a single-direction Theorem 2$'$ instance at a
\textbf{least-favorable mixture} of directions --- and the duality separates the two
obstructions exactly: randomization eliminates non-closure and can never
touch cross-leak.

\begin{nthm}{Lemma 3}[vector operating-point reduction]
Each $h$ has a joint
operating point $Q(h) := (q_0^j(h), q_1^j(h))_{j \leq k} \in [0,1]^{2k}$;
write $\mathrm{ROC}_k(H) := \{Q(h) : h \in H\}$. On the union family, the
deterministic inductive premium is
$$\Pi^{\mathrm{ind}}_L(H) \;=\; \sup_{Q \in \mathrm{ROC}_k(H)}\; \min_j \big[ A\,q_0^j - B\,q_1^j \big]$$
--- the max-min of the price functional over the joint ROC set, whose Pareto
frontier does not factorize per direction.
\end{nthm}

\begin{proof}
Per direction, Lemma 2's computation (worst $\lambda$ at
$\bar\varepsilon$); the adversary's choice of direction is the min.
\end{proof}

\textbf{Randomized play.} Let the agent draw its detector per task from a finite
lottery $\pi$ over $H$. The operating point of a lottery is the mixture of
its atoms' points, so lotteries realize exactly
$\mathrm{conv}\,\mathrm{ROC}_k(H)$ (Carath\'eodory: at most $2k+1$ atoms
suffice --- the optimal randomized detector is a \textbf{small menu with a coin}).
Write $\Pi^{\mathrm{rand}}_L(H)$ for the inductive premium over lotteries.

\begin{nthm}{Theorem M}[least-favorable mixture duality]
For any $H$ (constants
included), any $k$ ensemble directions, and every $L$:
$$\Pi^{\mathrm{rand}}_L(H) \;=\; \min_{w \in \Delta_k}\; \Pi^{2'}_L\!\Big(H;\; \sum_j w_j \nu_0^j,\; \sum_j w_j \nu_1^j\Big),$$
where $\Pi^{2'}_L(H; \nu_0, \nu_1) = \sup_h [A\,\nu_0(h{=}a) -
B\,\nu_1(h{=}a)]$ is Theorem 2$'$'s single-pair value. The minimizer $w^*$
is the \textbf{least-favorable direction mixture}; the optimal lottery is the
class-restricted Neyman--Pearson solution against the mixed ensembles
$(\bar\nu_0^{w^*}, \bar\nu_1^{w^*})$; and $w^*$ is supported on the
binding directions (complementary slackness).
\end{nthm}

\begin{proof}
$\Pi^{\mathrm{rand}}_L = \sup_{q \in \mathrm{conv}\,\mathrm{ROC}_k}
\min_j \langle c_j, q\rangle$ with $c_j$ the $j$-th price functional. The
objective is concave (min of linear) and unchanged by passing to the closed
convex hull; the adversary's set is finite; von Neumann/LP duality gives
$\sup_q \min_j = \min_{w \in \Delta_k} \sup_q \langle \sum_j w_j c_j,
q\rangle$, with the min attained on the compact simplex. Finally
$\sum_j w_j [A q_0^j(h) - B q_1^j(h)] = A\,\bar\nu_0^w(h{=}a) -
B\,\bar\nu_1^w(h{=}a)$ because operating points are linear in the measure ---
so the inner sup \emph{is} a Theorem 2$'$ instance.
\end{proof}

\begin{nthm}{Corollary M1}[the obstructions, exactly separated]
For every $L$,
$$\underbrace{\Pi^{\mathrm{tr}}_L - \Pi^{\mathrm{ind}}_L}_{\text{transduction gap}} \;=\; \underbrace{\big[\Pi^{\mathrm{tr}}_L - \Pi^{\mathrm{rand}}_L\big]}_{\text{irreducible cross-leak}} \;+\; \underbrace{\big[\Pi^{\mathrm{rand}}_L - \Pi^{\mathrm{ind}}_L\big]}_{\text{closure deficit}}.$$
\end{nthm}

The first bracket is the cross-leak against the least-favorable mixture --- it
survives any amount of randomization, because Theorem M already plays the
convex hull. The second is the convexity gap of $\mathrm{ROC}_k(H)$ at the
optimal price functional --- the \textbf{price of determinism}, and it is \emph{all} of
Prop 4's non-closure obstruction: \textbf{a coin buys back the gluing failure,
never the cross-leak.} Union-closed classes (Prop 5) have zero closure
deficit already; for them randomization is worthless, as the exact formula
showed.

\textbf{Worked example (XOR, the numbers).} Halfspaces on the XOR directions
(Appendix~\ref{app:deficit}, Prop 6), $L \to \infty$, in units of $A$: deterministic $0$, randomized
$1/2$ (the coin over the two corner halfspaces $\{x{+}y < \tfrac12\}$,
$\{x{+}y > \tfrac32\}$; dual check: at $w^* = (\tfrac12, \tfrac12)$ no
halfspace excludes both fatal points while capturing more than half the
mixed safe mass), transductive $1$. Both obstructions are worth exactly
$1/2$ here: the three values separate, and the decomposition of M1 is
witnessed with all brackets positive.

\begin{nthm}{Corollary M2}[$L \to \infty$]
$\Pi^{\mathrm{rand}}_L / A$ decreases to
the randomized joint coefficient, which under a closed joint ROC set equals
$\min_{w} \bar\sigma_0\big(H;\ \bar\nu_0^w, \bar\nu_1^w\big)$ --- the
zero-leak capture capacity at the least-favorable mixture. Without
closedness the limit sits between the $t{=}0$ and right-limit versions,
the same caveat as Appendix~\ref{app:thm2prime}'s closure remark; we do not pretend otherwise.
\end{nthm}

\textbf{Remarks.}
\begin{itemize}
\item The lineage is now verified (the project repository's literature notes,
  second round). The duality
  technology is fully classical: \citet{wald1945,wald1950} for minimax-equals-Bayes
  against a least favorable prior with randomized rules; \textbf{\citet{huberstrassen1973}}
  for the exact collapse move --- composite minimax testing reduces
  to one NP test against a least favorable pair; \citet{fillatre2017} for the
  finite-LP/dual-LP computation. Even the coin mechanism is in print:
  randomized rules realize the ROC convex hull (NP lemma, \citealt[Thm 3.2.1]{lehmannromano2005};
  \citealt{provost2001}; \citealt[\S V-D]{fauss2021} on point-mass LRs
  forcing randomization). We cite all of it and claim only the application ---
  and must distinguish \citet{managoli2025},
  who already pair robust testing with abstention (asymptotic exponents,
  unrestricted detectors): M's claim sits at the restricted-class premium
  game, the closure-deficit identification, and the obstruction split.
\item Theorem R is now the \emph{constructive} counterpart of M: blind randomization
  (route without looking) is the $\rho = 1 - 1/k$ corner of the router bound,
  and M says the best blind play is not gating but the NP solution against
  $w^*$; a router with per-task information interpolates between M's value
  and the transductive one.
\item What M does \textbf{not} deliver: the deterministic value itself. The max-min
  over a non-convex $\mathrm{ROC}_k(H)$ (Lemma 3) is exact but not, in
  general, simplifiable --- computing it is plausibly hard (the nearby
  classifier-rejector hardness of \citealt{mozannar2023} suggests so, but we
  have not proved it), and the closure deficit is exactly what it costs. That
  is now the honest residual.
\end{itemize}

\section{Corollary 3 --- the growth form}\label{app:growth}

\begin{nthm}{Corollary 3}
Under multiplicative dynamics (wealth factor $1+r$ per task:
$r = \tilde g$ on safe acts, $r = -\tilde L$ on missed fatals, $r = -\tilde p$
on escalations, i.i.d.\ tasks), with a confounded detector class: every
within-defect policy with positive specialization premium has, on some
$P \in \mathcal{P}$, expected log-growth $\to -\infty$ as $\tilde L \to 1$;
every within-defect policy with bounded log-growth has premium $0$.
\end{nthm}

\begin{proof}
Positive premium requires positive act-rate $q$ on the confounded
family (Theorems 1, 2), and $\mathbb{E}_{P_{\bar\varepsilon}}[\log(1+r)]$
contains the term $\bar\varepsilon\, q \log(1 - \tilde L) \to -\infty$.
Policies avoiding this have $q = 0$, hence premium $0$ by the same theorems.
\end{proof}

\textbf{Survival and specialization gain cannot coexist inside the defect.} This
is Proposition 3 of the project's formalization notes with its missing half: there, ruin
forces $\mu = 0$; here, inside the defect, $\mu = 0$ forces $q_0 = 0$
(Lemma 1) --- no gain. The two halves together are the theorem the founding
statement needed.

\section{Theorem A --- the achievability half (the converse's partner)}\label{app:thmA}

The converse (Appendices \ref{app:thm1}--\ref{app:thm2prime}) says a confounded detector earns zero premium. The
partner statement: a detector placed \emph{outside} the defect earns a premium
bounded below by an explicit constant, so removal is not merely possible but
\emph{profitable} --- closing the ``if'' direction of the iff. Where Appendices \ref{app:thm1}--\ref{app:thm2prime} fixed the
detector class and varied $P$ adversarially, here we fix a good detector and
lower-bound its worst-case premium over the class.

\textbf{Hypotheses.}
\begin{itemize}
\item \textbf{Structured class with a competence floor.}
  $\mathcal{P} \subseteq \{P : P(E) \leq \bar\varepsilon,\; P(C) \geq \underline{c}\}$
  with $\underline{c} + \bar\varepsilon \leq 1$. The floor $\underline c > 0$
  is necessary: if the world may contain no in-competence mass, no policy can
  guarantee positive premium (there is nothing to specialize \emph{on}).
\item \textbf{Out-of-defect detector.} There is a detector $d$ achieving, \emph{uniformly
  over $\mathcal{P}$}, miss rate $\sup_P P(d{=}a \mid E) \leq \delta$ and
  false-alarm rate $\sup_P P(d{=}e \mid C) \leq \alpha_0$, at rent $c_d$, and
  escalating the benign region ($P(d{=}a \mid N) = 0$ --- the conservative
  choice; acting on benign only helps, by $(p-\ell) \geq 0$ per unit). ``Out of
  the defect'' is exactly what makes both bounds simultaneously attainable:
  Lemma 1 forbids it \emph{inside} an observation defect (there capture rate $=$
  miss rate), and Appendix~\ref{app:thmE} \emph{constructs} such a $d$ for
  structured $\mathcal{P}$ (declared covers with finite-VC detector classes,
  Theorem~E$'$; Mondrian/group-conditional conformal constructions
  \citep{vovk2005} cover the same ground distribution-free). A inherits that
  structural restriction --- no more, no
  less.
\end{itemize}

\textbf{The premium identity.} Against the always-escalate baseline $B_0 = -p$,
for any detector and any $P$ (writing $\alpha_P, \mu_P$ for the realized
rates, benign escalated):
$$\Pi_P(d) \;=\; P(C)\,(1-\alpha_P)\,(g+p) \;-\; P(E)\,\mu_P\,(L-p) \;-\; c_d.$$

\begin{proof}[Derivation]
Add $p = p\,[P(C)+P(E)+P(N)]$ to $G_P(d)$ region by region. On
$C$: $P(C)[(1-\alpha_P)g - \alpha_P p] + pP(C) = P(C)(1-\alpha_P)(g+p)$. On
$E$: $-P(E)p - P(E)\mu_P(L-p) + pP(E) = -P(E)\mu_P(L-p)$ (baseline exactly
cancels the escalate cost on $E$). On $N$ with benign escalated: $-pP(N) +
pP(N) = 0$. Subtract rent $c_d$.
\end{proof}

Three terms, each a per-unit price: \textbf{captured competence} earns $(g+p)$ per
unit (gain $g$ where the baseline paid $p$); \textbf{leaked fatal} costs $(L-p)$
per unit (loss $L$ where the baseline paid $p$); \textbf{rent} is flat.

\begin{nthm}{Theorem A}
Under the hypotheses,
$$\inf_{P \in \mathcal{P}} \Pi_P(d) \;\geq\; \Pi^* \;:=\; \underline{c}\,(1-\alpha_0)\,(g+p) \;-\; \bar\varepsilon\,\delta\,(L-p) \;-\; c_d,$$
and the guaranteed premium is positive ($\Pi^* > 0$) --- hence removal
profitable --- iff the
\textbf{advantage condition} holds:
$$\underline{c}\,(1-\alpha_0)\,(g+p) \;>\; \bar\varepsilon\,\delta\,(L-p) + c_d.$$
\end{nthm}

\begin{proof}
The identity gives $\Pi_P(d) = P(C)(1-\alpha_P)(g+p) - P(E)\mu_P(L-p)
- c_d$, which is nondecreasing in $P(C)$ and nonincreasing in each of
$\alpha_P, \mu_P, P(E)$. Apply the uniform caps $\alpha_P \leq \alpha_0$,
$\mu_P \leq \delta$ and the class bounds $P(C) \geq \underline c$, $P(E) \leq
\bar\varepsilon$ --- all valid for every $P \in \mathcal{P}$ --- to get
$\Pi_P(d) \geq \Pi^*$ pointwise, hence for the infimum.
\end{proof}

\textbf{This is Proposition 1 / the advantage condition (\S\ref{sec:advantage} of the main text),
re-derived as the achievability threshold} with the detector's real
operating characteristics substituted for the abstract ``price of
compensation'': the escalation cost enters as the $(1-\alpha_0)$ haircut on
captured competence, and the missed-fatal cost as $\bar\varepsilon\delta(L-p)$.

\subsection*{Remarks --- where A meets B}

\textbf{1. The $L$-dependence, and the exact meeting with the converse.} $\Pi^*$
carries $L$ only through the leakage $\bar\varepsilon\delta(L-p)$. With
$\delta > 0$ \emph{fixed}, $\Pi^* \to -\infty$ as $L \to \infty$: \textbf{a detector of
any fixed positive miss rate is eventually ruined by a severe enough fatal
event.} This is not a weakness of the proof --- it is A shaking hands with
Conjecture B, which drove $G \to -\infty$ precisely when $\mu$ stayed bounded
away from $0$. The two theorems partition the $(\delta, L)$ plane along the
same curve. Profitable removal survives arbitrary $L$ only if the miss rate
shrinks with severity,
$$\delta(L) \;<\; \frac{\underline c(1-\alpha_0)(g+p) - c_d}{\bar\varepsilon(L-p)} \;=\; O\!\big(1/L\big).$$

\textbf{2. Two independent routes to a bound uniform in $L$.} Either
(i) \textbf{sharpen the detector}, $\delta = O(1/L)$, so leakage stays $\leq
\bar\varepsilon\kappa$ for a constant $\kappa$; or (ii) \textbf{cap the blast
radius} --- a spend ceiling / bounded compensation making realized loss
$\leq \bar L$ regardless of the true $L$ (formalization notes \S 2, point 2), after
which any fixed $\delta$ satisfying the advantage condition at $\bar L$ works.
The architecture uses both; either alone suffices for a constant lower bound.

\textbf{3. Route (i) is cheap --- the tie to detector economics.} By the
Chernoff--Stein hypothesis-testing bound \citep[Thm.~11.8.3]{coverthomas2006},
miss rate
$\delta$ costs rent $c_d \sim \log(1/\delta)$, so $\delta = O(1/L)$ costs only
$c_d \sim \log L$. Leakage is held constant at the price of \emph{logarithmic}
rent --- the detector's cost grows exponentially slower than the severity it
insures against. Profitability then holds up to a finite but \textbf{exponentially
large} ceiling $L \leq \exp\big(O(\text{margin})\big)$, where the margin is
the specialization premium net of constant leakage. Cheap detection is what
makes the whole architecture affordable, quantified.

\textbf{4. Graceful degradation recovers the converse's witness.} At the ideal
operating point $\alpha_0 = \delta = 0$ and zero rent $c_d = 0$ (perfect free
detector) with the fullest
class $\underline c = 1-\bar\varepsilon$, $\Pi^* = (1-\bar\varepsilon)(g+p)$
--- \textbf{exactly the separation witness of Theorem 1(3)} and the outside-defect
value against which the converse measured the collapse. A's bound degrades
continuously from that ideal as $(\alpha_0, \delta, c_d)$ rise, so Theorems 1$'$
and A are two ends of one continuum: $\sigma_0$ measures how much of the
premium a \emph{confined} detector keeps; $(\alpha_0, \delta)$ measure how much a
\emph{free but imperfect} detector keeps.

\subsection*{Corollary A+B --- the full iff}

Combining Theorem A with the converse (Appendices \ref{app:thm1}--\ref{app:thm2prime}), over structured classes and
with either $L$ capped or $\delta = O(1/L)$:

\begin{quote}\itshape
A defect is \textbf{profitably removable} iff a detector achieving small
miss and false-alarm rates exists \textbf{outside} it --- equivalently, iff the
detector-relevant distinction survives the restriction (zero-leak capture
capacity $\bar\sigma_0 > 0$, one coefficient for both defect types by
Theorem 2$'$) \textbf{and} the advantage condition holds.
\end{quote}

Necessity is B (inside the defect the distinction is gone, premium $= 0$);
sufficiency is A (outside it, the explicit $\Pi^* > 0$). Placement supplies
the \emph{possibility}; the advantage condition supplies the \emph{profit}. This is the
theorem the founding statement asked for, in both directions.

\section{Theorem E --- end-to-end achievability: the detector learned, not assumed}\label{app:thmE}

Theorem A takes the uniform rate caps $(\alpha_0, \delta)$ as given. Here the
detector is \emph{selected from data}, and the caps are replaced by certificates.
The concentration is standard and we use it as such (textbook constants, not
optimized); the content is where the union bound lands and what the
certificate costs as severity grows.

\textbf{Setup (declared cover; the repaired-C structure made concrete).} The
agent declares a finite \textbf{cover}: safe-cell conditionals
$\nu_0^1, \dots, \nu_0^r$ (supported in $C$) and fatal-cell conditionals
$\nu_1^1, \dots, \nu_1^m$ (supported in $E$) --- fixed measures, a
Mondrian/group-conditional structure (cf.\ Mondrian conformal prediction,
\citealt{vovk2005}) in its simplest
form. The uncertainty class $\mathcal{P}$ reweights cells arbitrarily subject
to the budgets $P(E) \leq \bar\varepsilon$, $P(C) \geq \underline c$; within
cells, conditionals are fixed. Worst-case rates then reduce to cell maxima:
$\mu(d) := \max_i \nu_1^i(d{=}a)$, $\alpha(d) := \max_l \nu_0^l(d{=}e)$, and
Theorem A's bound holds with these: $\inf_P \Pi_P(d) \geq \Pi^{\mathrm{lb}}(d)
:= \underline c\,(1-\alpha(d))(g+p) - \bar\varepsilon\,\mu(d)(L-p) - c_d$.

The detector class $D$ (VC dimension $v$, outside the defect --- that is what
$D$'s freedom means) is \textbf{gated by the cover}: every $d \in D$ escalates
off the covered cells. Data: $n$ i.i.d.\ samples from each cell conditional
(stratified --- test the candidate detector against known-fatal history per
category).

\begin{nthm}{Lemma 4}[uniform cell concentration; standard]
With probability
$\geq 1-\theta$, simultaneously for all $d \in D$ and all $m + r$ cells, the
empirical rates satisfy $|\hat\mu_i(d) - \mu_i(d)| \leq \epsilon_n$ and
$|\hat\alpha_l(d) - \alpha_l(d)| \leq \epsilon_n$ with
$$\epsilon_n = O\!\Big(\sqrt{\tfrac{v \log n + \log((m+r)/\theta)}{n}}\Big).$$
\end{nthm}

\begin{proof}
Two-sided VC uniform convergence per cell \citep{vapnik1971}, union bound over
the $m+r$ cells --- the promised ``union bound over
$\mathcal{P}$'s structure,'' costing only $\log(m+r)$.
\end{proof}

\begin{nthm}{Theorem E}[agnostic]
Let $\hat d$ maximize the empirical bound
$\hat\Pi^{\mathrm{lb}}$. On Lemma 4's event,
$$\inf_{P \in \mathcal{P}} \Pi_P(\hat d) \;\geq\; \sup_{d \in D} \Pi^{\mathrm{lb}}(d) \;-\; 2\kappa\,\epsilon_n, \qquad \kappa := \underline c\,(g+p) + \bar\varepsilon\,(L-p).$$
\end{nthm}

\begin{proof}
$\Pi^{\mathrm{lb}}$ is $\kappa$-Lipschitz in the sup-norm of the
rate vector (cell-max is 1-Lipschitz), so $|\Pi^{\mathrm{lb}} -
\hat\Pi^{\mathrm{lb}}| \leq \kappa\epsilon_n$ uniformly; sandwich the ERM.
\end{proof}

The penalty is priced by the same $\kappa \approx B$ for large $L$: under the
square-root rate, holding the estimation penalty constant costs $n \sim L^2$
samples. The agnostic training bill is \textbf{quadratic in severity} --- too
expensive. The rare-event form fixes this:

\begin{nthm}{Theorem E$'$}[rare-event certificate --- the linear bill]
Suppose the fatal
side is realizable: some $d^* \in D$ has $\mu_i(d^*) = 0$ on every declared
fatal cell, with $\alpha(d^*) \leq \alpha^*$. Select $\hat d$ among
$\{d : \hat\mu_i(d) = 0\ \forall i\}$ (nonempty \textbf{surely} --- a true-zero
detector cannot miss a sample) minimizing $\max_l \hat\alpha_l$. With
probability $\geq 1-\theta$:
$$\mu(\hat d) \;\leq\; \epsilon^{01}_n = O\!\Big(\tfrac{v \log n + \log(m/\theta)}{n}\Big), \qquad \alpha(\hat d) \;\leq\; \alpha^* + 2\epsilon_n,$$
(one-sided realizable version-space bound --- \citealt[Thm 2.1]{blumer1989} --- union over
fatal cells; two-sided only on the safe side; E$'$'s problem statement is
KKM positive-reliable learning and its setup is NP classification at
$\alpha = 0$, see Appendix~\ref{app:positioning}), hence
$$\inf_{P} \Pi_P(\hat d) \;\geq\; \underline c\,(1 - \alpha^* - 2\epsilon_n)(g+p) \;-\; \bar\varepsilon\,\epsilon^{01}_n\,(L-p) \;-\; c_d.$$
Holding the fatal-side penalty below a constant requires
$\epsilon^{01}_n = O(1/L)$, i.e.\ $n = O\big(L \cdot (v\log L + \log(m/\theta))\big)$
per fatal cell --- \textbf{linear in severity}, because certifying a near-zero rate
is a one-sided rare-event problem, not a mean estimation. The safe side needs
only $O(1/\sqrt{n})$, independent of $L$. \qquad $\square$
\end{nthm}

\textbf{Lifecycle economics.} Sustaining severity $L$ end-to-end now has two
bills: a \textbf{one-time training bill} of $O(L \cdot v \log L)$ labeled fatal
examples per declared category (amortized over deployment), and the
\textbf{per-period rent} $c_d = O(\log L)$ from Appendix~\ref{app:thmA}'s Remark 3. Amortized, both
grow slower than the per-event loss they insure against, so \textbf{the advantage
condition survives learning} whenever it held with known rates, up to the
explicit slack above. One economic observation rides along: the scarce input
is \emph{labeled catastrophes} --- precisely what a well-run system stops producing
(it escalates them). The detector's training data must come from history,
simulation, or other agents' failures; in market terms, labeled fatal
examples are a tradable good, and Theorem E$'$ prices demand for them at
$O(L)$ per category.

\textbf{Coverage: fail-safe against incompleteness, exposed to mis-declaration.}
Because $D$ is gated by the cover, a fatal category \emph{outside} the declared
cells is escalated by default --- incompleteness costs foregone premium (an
uncovered safe cell earns nothing), never uncertified fatal exposure. The
certificate's true residual risk is \textbf{mis-declaration}: fatal mass inside a
declared-\emph{safe} conditional's support, where the gate admits and the samples
mislead. That is model misspecification, not sample complexity, and it is
mitigated structurally (the blast-radius cap on $L$), not statistically ---
the formalization notes' honesty rule, surviving to the end of the pipeline.
This also discharges Theorem A's delegation: existence of the out-of-defect
detector is no longer assumed but reduced to realizability of $D$ on the
declared cells, with everything else certified from data.

\section{Positioning}\label{app:positioning}

\begin{itemize}
\item \textbf{vs.\ \citet{fang2022}:} their impossibility is statistical
  (non-learnability of OOD detection in restricted spaces); ours is
  decision-theoretic (value collapse for an actor--detector pair sharing one
  restriction), with the adversarial-distribution move imported as
  technology, as planned.
\item \textbf{vs.\ \citet{goldwasser2020}:} not a contradiction but a corner of the
  theory --- their escape is Theorem 2's minimax gap, and Theorem 1(2) marks
  its boundary: transduction rescues capacity defects, never observation
  defects.
\item \textbf{vs.\ Littlewood--Rushby / common-cause failure:} Lemma 1 is the
  common-cause principle upgraded from a correlation model to an identity
  ($q_0 = q_1$), with the quantitative wedge $\tau$ recovering the graded
  version \citep{littlewood2012}.
\item \textbf{vs.\ Chow:} within the defect, optimal play \emph{is} Chow's rule on the
  coarsened posterior (Theorem 1$'$ is a Neyman--Pearson problem) --- and that is
  precisely why optimal within-defect play cannot help. The failure is not in
  the decision rule but in what reaches it.
\item \textbf{vs.\ least-favorable priors (verified):} Theorem M's structure is the
  classical minimax-testing duality --- \citet{wald1945,wald1950}, \citet{huber1965},
  \citet{huberstrassen1973} for the collapse-to-one-NP-test move, \citet{fillatre2017}
  for the finite-LP form; randomization-realizes-the-ROC-hull is
  NP-lemma classical (\citealt[Thm 3.2.1]{lehmannromano2005}; \citealt{provost2001};
  \citealt[\S V-D]{fauss2021}). We claim the application only, and distinguish
  \citet{managoli2025} --- robust testing \emph{with} abstention
  exists as of 2025, at asymptotic exponents over unrestricted detectors;
  the restricted-class premium game and the obstruction split are ours.
  Affirmative evidence the classical robust-detection canon skipped the
  reject option: the Fau\ss--Zoubir--Poor survey contains no abstention action.
\item \textbf{vs.\ NP classification and reliable learning (verified):} Theorem E is
  standard technology (\citealt{vapnik1971}; \citealt[Thm 2.1]{blumer1989} for the version-space
  bound; textbooks \citealt{anthonybartlett1999}, \citealt{shalevshwartz2014}). Theorem E$'$'s
  \emph{problem} is Kalai--Kanade--Mansour's positive-reliable learning \citep{kalai2012} and its
  \emph{setup} is Neyman--Pearson classification at $\alpha = 0$ (\citealt{cannon2002};
  \citealt{scottnowak2005}; \citealt{rigollet2011}) --- cite both frames. Neither
  contains the zero-empirical-miss selection with the one-sided fast rate
  (KKM's reduction pays agnostic $\varepsilon^2$ rates; all NP papers run
  two-sided $\sqrt{\,}$ at $\alpha > 0$), nor the linear-vs-quadratic-in-$L$
  training bill --- which we present as an elementary corollary of the tight
  realizable/agnostic $1/\varepsilon$ vs $1/\varepsilon^2$ dichotomy (EHKV
  lower bound: \citealt{ehrenfeucht1989}), not a new phenomenon. Must engage: \citet{casacuberta2025},
  group-wise reliable abstention at agnostic rates --- owns
  the group-wise frame; lacks zero-miss certification, reweighting
  robustness, and the severity asymmetry.
\item \textbf{The slogan, final form:} \emph{a defect is profitably removable exactly to
  the extent that the detector's distinction survives outside it} ---
  coefficient $\sigma_0$, not a yes/no.
\end{itemize}

\section{Open}\label{app:open}

\begin{enumerate}
\item \textbf{Iterated adversary} (research backlog \#3, second half): the adversary here
   chooses $P$ once; the repeated game with a learning detector is untouched.
\item \textbf{Beyond mixture families:} the theorems quantify over
   $\mathcal{F}(\nu_0,\nu_1)$ and unions thereof; characterizing collapse for
   arbitrary $\mathcal{P}$ (not built from confusable ensemble pairs) is open,
   though any $\mathcal{P}$ \emph{containing} such a family inherits the collapse.
\item \textbf{The deterministic max-min.} Theorem M resolves the mixed-obstruction
   deficit for randomized play; the \emph{deterministic} value (Lemma 3's max-min
   over the non-convex joint ROC set) is exact but unsimplified, and its
   computational complexity is unproved --- the nearby classifier-rejector
   hardness \citep{mozannar2023} suggests NP-hardness for halfspace-type
   classes, but we have not shown it. Equivalently open: bounding the closure
   deficit by a structural property of $H$ (something Shapley--Folkman-shaped
   as $k$ grows).
\item \textbf{The iterated router.} Theorem R's recursion is one level deep. A
   hierarchy of directions (directions about directions) would iterate it;
   whether the total rent telescopes (each level $O(\log L)$) or compounds
   is open, and bears directly on the research backlog's \#5 fleet economics.
\end{enumerate}

\bibliographystyle{plainnat}
\bibliography{references}

\end{document}